\documentclass[prl,twocolumn,showpacs,preprintnumbers,amsmath,amssymb]{revtex4}

\usepackage[dvips]{graphicx}
\usepackage{dcolumn}
\usepackage{bm}
\usepackage{times}
\usepackage{mathptmx}

\begin{document}

\title{Two-Dimensional Propagation of a Photoinduced Spin Wave Packet}

\author{Yuki~Terui$^1$}
\author{Takuya~Satoh$^{1,2}$}
\email{tsatoh@iis.u-tokyo.ac.jp}
\author{Rai~Moriya$^1$}
\author{B. A. Ivanov$^{1,3}$}
\author{Kazuya~Ando$^4$}
\author{Eiji~Saitoh$^{4,5,6}$}
\author{Tsutomu~Shimura$^1$}
\author{Kazuo~Kuroda$^1$}

\affiliation{$^1$Institute of Industrial Science, The University of Tokyo, Tokyo 153-8505, Japan}
\affiliation{$^2$PRESTO, Japan Science and Technology Agency, Sanbancho, Tokyo 102-0075, Japan}
\affiliation{$^3$Institute of Magnetism, Vernadskii Avenue 36B, 03142 Kiev, Ukraine}
\affiliation{$^4$Institute for Materials Research, Tohoku University, Sendai 980-8577, Japan}
\affiliation{$^5$CREST, Japan Science and Technology Agency, Sanbancho, Tokyo 102-0075, Japan}
\affiliation{$^6$The Advanced Science Research Center, Japan Atomic Agency, Tokai 319-1195, Japan}

\begin{abstract}
We report the two-dimensional propagation of photoinduced spin wave packets 
in Bi-doped rare-earth iron garnet. 
Spin waves were excited nonthermally and impulsively by a circularly polarized light pulse 
via the inverse Faraday effect. 
Space- and time-resolved spin waves were detected with a magneto-optical pump-probe technique. 
We investigated propagation in two directions, parallel and perpendicular to the 
magnetic field. Backward volume magnetostatic waves (BVMSWs) were detected in both directions. 
The frequency of BVMSWs depends on the propagation direction. The experimental results agreed well 
with the dispersion relation of BVMSWs. 
\end{abstract}


\date{
\today}

\maketitle

Magnetization dynamics has attracted a great deal of interest in recent years. In particular, the propagation 
characteristics of spin waves have been extensively studied because of their importance as 
the basis of magnonics, which aims to use packets of spin waves instead of electric currents as the 
information transmitter \cite{Serga,Krug}. 
Magnonics shows great promise to improve solid-state devices 
in combination with conventional electronics \cite{Saitoh}. 
When spins are coupled mainly by the magnetic dipolar 
interaction rather than the exchange interaction, the spin waves are called 
magnetostatic waves (MSWs) \cite{Stancil}. 
Traditionally, MSWs have been excited by a microwave field emitted from a microstrip antenna. 
Time-resolved MSWs have been examined in many ways, for example, via the inductive method \cite{Covington,Wu}, 
the magneto-optical Kerr effect \cite{Silva,Park,Barman,Liu,Perz}, and Brillouin light scattering \cite{Serga2,Demo}. 
Their nonlinear effects \cite{Demidov} and mode quantization \cite{Demidov2} have also been investigated. 

Analysis of spin waves excited by an ultrashort light pulse has already been reported \cite{Lenk}. 
In that study, the spin waves were excited by a heat-induced change in the sample's anisotropy. 
Nonthermal excitation is thus desirable from the viewpoint of exciting coherent spin waves 
and simplifying the analysis. 
Recently, a novel method of magnetization control has been reported. 
The method employs the inverse Faraday effect (IFE) \cite{Kimel}, where a circularly polarized light pulse 
nonthermally generates an effective magnetic field. 
IFE has been described as impulsive stimulated Raman scattering \cite{Hansteen}. 
Thus, ultrafast magnetic switching is expected via IFE. 
It has been reported that IFE occurs in diverse magnetic 
materials \cite{Kimel,Hansteen,Stanciu,Kalash,Satoh}. However, in these reports, discussion has been limited to local spin oscillations only. 
Non-local spin dynamics, i.e., \textit{propagating spin waves}, excited by this impulsive excitation method has 
never been observed. 
By using this method, two-dimensional propagation 
can be observed in a non-contact manner, and broadband excitation can be realized, unlike resonant excitation with a microwave field. 

Yttrium iron garnet (YIG) has often been used as a sample in magnonics because of its intrinsically low magnetic 
damping \cite{Stancil}. In an experiment involving photoinduced spin waves, Bi-doped rare-earth iron garnet was found to be
more suitable than YIG due to its high magneto-optical susceptibility \cite{Lacklison}. 
By doping the sample with bismuth and magnetic rare-earth ions, the Gilbert damping coefficient increases. 
Even so, MSWs can propagate over distances of hundreds of micrometers, and they can be spatially resolved. 

\begin{figure}[b]
\includegraphics[height=3.5cm,clip,keepaspectratio]{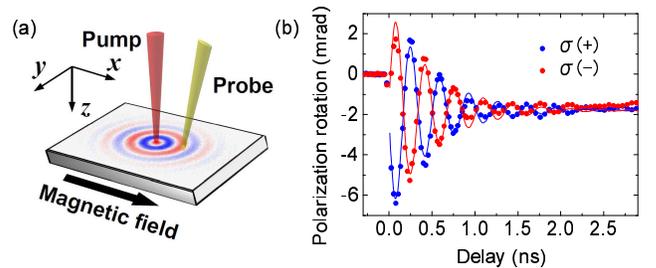}
\caption{(Color online) (a) The experimental configuration. The sample surface was in the $x$-$y$ plane. 
The magnetic field was applied in the $x$ direction. 
(b) Time-resolved magneto-optical Faraday rotation waveforms when pump spot and probe spot
overlapped spatially. 
$\sigma(+,-)$ represent pump helicities. 
Solid lines represent fitting results.}
\label{FIG. 1.}
\end{figure}

Here we report the two-dimensional propagation of a spin wave packet excited via IFE 
in Bi-doped rare-earth iron garnet. 
Propagation in two directions, parallel and perpendicular to the magnetic field, was investigated by an 
all-optical pump-probe experiment. 

The composition of our sample was $\rm{Gd}_{4/3}\rm{Yb}_{2/3}\rm{Bi}\rm{Fe}_5\rm{O}_{12}$, 
and the thickness was 110~$\mu$m. The unit cell is cubic 
with three types of interstices, 16a, 24d, and 24c sites \cite{Aulock}. 
$\rm{Fe}^{3+}$ ions fill the 16a and 24d sites, and rare-earth ions fill the 24c sites. 
The magnetic moments of $\rm{Fe}^{3+}$ ions in different sites are antiferromagnetically coupled. 
The magnetic moments of rare-earth ions have much weaker antiferromagnetic coupling 
with the net $\rm{Fe}^{3+}$ moment. 
Thus, saturation magnetization $(4{\pi}M_s$ = 1136 G$)$ is reduced compared with 
the typical value for YIG $(4{\pi}M_s$ = 1956 G$)$ \cite{Saitoh}. 

\begin{figure}[t]
\includegraphics[height=4cm,clip,keepaspectratio]{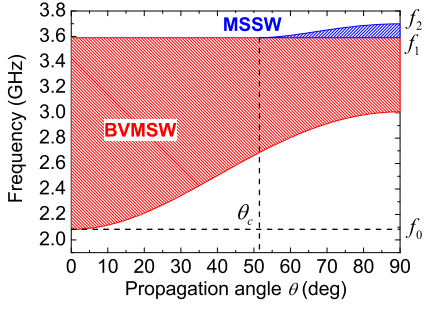}
\caption{(Color online) The BVMSW band and MSSW band versus in-plane propagation angle $\theta$ 
when $H_0 = 1000$ Oe and $H_u = 450$ Oe. 
The lower limit of the BVMSW band and the upper limit of the MSSW band moved up 
as $\theta$ increased. MSSW emerged when $\theta > \theta_c$. } 
\label{FIG. 2.}
\end{figure}

Our sample had uniaxial anisotropy. The $z$ axis was the easy axis, and the $x$-$y$ plane was the hard plane (See Fig. 1(a)). 
To evaluate the uniaxial anisotropy field, the uniform precession frequency $f_1$ was 
determined to be 3.6 GHz under an in-plane external field of 1000 Oe in a magnetic resonance experiment. 
The value was compared by using
\begin{equation}
f_1 = {\gamma}\sqrt{H_0\left(H_0-H_u+4{\pi}M_{\rm{eff}}\right)},
\end{equation}
where $H_0$ is the applied magnetic field, and $H_u$ is the uniaxial anisotropy field to the $z$ axis. 
Here, $4{\pi}M_{\rm{eff}}$ is defined as $\left(N_z-N_x\right)4{\pi}M_s$, 
$N_{x(z)}$ is the demagnetizing factor along the $x(z)$ axis, and 
$\gamma$ = 2.8 $\rm{MHz}/\rm{Oe}$ is the electron gyromagnetic ratio. 
Then, $H_u$ was determined to be 450 Oe. 
From the FWHM of the resonance peak, we also determined the Gilbert damping coefficient to be $\alpha = 0.02$, 
which is about $10^2$ times higher than the typical value for YIG $(\sim 10^{-4})$. 
This difference was attributed to an increase in indirect 
coupling between the $\rm{Fe}^{3+}$ sublattice and the phonon lattice via rare-earth ions \cite{Sparks}. 

We employed the magneto-optical Faraday effect to measure space- and time-resolved MSWs. 
The experimental configuration is shown in Fig. 1(a). 
Linearly polarized pulses from a Ti:sapphire 
laser with a wavelength of 792 nm and a pulse width of 120 fs were used as 
the probe. Circularly polarized pulses with a wavelength of 1400 nm, generated by an 
optical parametric amplifier, were used as the pump. 
The pump beam was vertically incident on the surface of the 
sample, whereas the probe beam was incident at 7$^\circ$. The pump and probe beams were 
focused on the sample to spot diameters of about 50 and 40 $\mu$m, respectively. 
The spot sizes were sufficiently small to resolve MSWs spatially, as will be confirmed below. 
The photon energy of the pump pulses (0.89 eV) was below the lowest \textit{d-d} transition in 
$\rm{Fe}^{3+}$ ions (1.4 eV) \cite{Scott}, leading to a virtual excitation. 
The pump fluence was 400 mJ/cm$^2$. 
In our sample, the refractive index was 2.4, and the absorption coefficient was 0.3 $\rm{cm^{-1}}$ for the pump wavelength. 
Therefore, absorbed pump fluence was 1 mJ/cm$^2$, which enabled almost uniform excitation across the entire thickness. 
An in-plane magnetic field of 1000 Oe was applied to saturate the magnetization in the $x$ direction. 
The pump spot was scanned in the directions parallel ($x$ axis) 
and perpendicular ($y$ axis) to the magnetic field at intervals of 20 $\mu$m, whereas the position of the probe spot was fixed. 
Polarization rotation of the probe light was recorded for each pump position. 
When the pump spot was scanned in the $x$ direction, we call it the parallel geometry. 
When the pump spot was scanned in the $y$ direction, we call it the perpendicular geometry. 
All measurements were performed at room temperature, which is below the Curie temperature of our sample ($T_c$ = 573 K). 

\begin{figure}[t]
\includegraphics[height=4cm,clip,keepaspectratio]{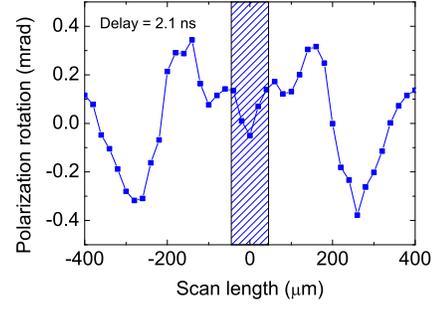}
\caption{(Color online) A representative spatial map of MSWs when scanning parallel to the magnetic field. 
The delay was fixed to 2.1 ns. The sum of the spot diameters of the pump and probe corresponds to the width 
of the shaded region.} 
\label{FIG. 3.}
\end{figure}

Figure 1(b) shows polarization rotation versus the time delay between the probe pulse and the pump pulse 
when the two spots overlapped spatially. 
The phase of spin oscillation shifted by $\pi$ rad when the helicity of the pump light reversed. 
This is clear evidence of IFE, where the direction of the magnetic field is determined by the pump helicity. 
The damped oscillations were fitted by $A\exp(-t/\tau)\sin(2{\pi}ft)-B$. 
The fitting results yielded $A = 5.3$ mrad, $B = 1.8$ mrad, $\tau = 0.52$ ns, and $f = 2.9$ GHz. 
Because the direction of the effective magnetic field pulse generated via IFE was perpendicular 
to the sample surface, the oscillations showed sinusoidal-dependence. 
In a sample having a Gilbert damping coefficient $\alpha$, the spin precession is damped in the direction of 
the magnetic field on a time scale of $1/\alpha\omega$ \cite{Stancil}. In our case, $1/\alpha\omega = 2.7$ ns is considerably longer 
than the value of $\tau$ determined by the fitting. 
This suggests that the energy and angular momentum propagated away from the pump spot in the form of spin waves. 
The relatively long-tailed (over 3 ns) decrease of polarization rotation, represented by $B$, was independent of the
pump helicity and was 
attributed to a heat-induced change in anisotropy. Nevertheless, IFE clearly dominated the excitation of 
spin oscillations within 3 ns. 
This thermal effect is subtracted in later analyses. 
Hereafter, the pump helicity was fixed to $\sigma(+)$. 
\begin{figure}[t]
\includegraphics[height=4cm,clip,keepaspectratio]{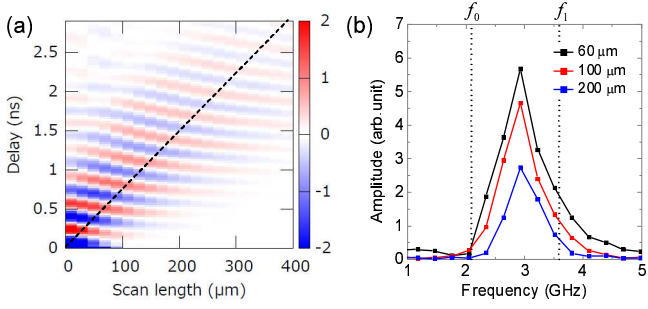}
\caption{(Color online) Spin waves when scanning parallel to the magnetic field. 
(a) A spatiotemporal map of spin waves. 
The color bar represents polarization rotation of the transmitted probe light in units of mrad. 
The dotted line represents calculation results of group velocity based on Eq. (4). 
(b) Dependence of Fourier spectrum of MSWs on scan length. 
The dotted lines respectively represent calculated upper and lower limits of the BVMSW band.}
\label{FIG. 4.}
\end{figure}

In an in-plane magnetized plate, two types of MSWs exist: 
backward volume magnetostatic waves (BVMSWs) and magnetostatic surface waves (MSSWs) \cite{Damon}. 
Hereafter, we discuss MSWs in a uniaxial anisotropic plate with perpendicular easy axis. 
In-plane propagation angle $\theta$ is defined as the angle between the direction of 
the magnetic field ($x$ axis) and the in-plane spin-wavevector. 
BVMSWs propagate at any angle to the magnetic field, whereas 
MSSWs propagate at angles greater than a critical angle $\theta_c = \sin^{-1}\left(\sqrt{{\gamma}H_0/f_1}\right)$ 
to the magnetic field \cite{Patton}. 
These MSWs exist in a finite range of frequencies, which are called the BVMSW band and the MSSW band, respectively. 
Figure 2 shows the frequency range of the bands versus in-plane propagation angle $\theta$ 
for $H_0 = 1000$ Oe and $H_u = 450$ Oe. 
When $\theta = 0^\circ$, the BVMSW band exists in a range $f_0 < f < f_1$. 
Here, $f_0$ is defined as
\begin{equation}
f_0 = {\gamma}\sqrt{H_0\left(H_0-H_u\right)}. 
\end{equation}
The lower limit of the BVMSW band moves up from $f_0$ to 
${\gamma}\sqrt{\left(H_0-H_u\right)\left(H_0+4{\pi}M_{\rm{eff}}\right)}$ 
as the propagation angle $\theta$ increases from 0$^\circ$ to 90$^\circ$. 
MSSWs emerge when $\theta > \theta_c$ = 51$^\circ$, and the upper limit of the MSSW band moves up from 
$f_1$ to $f_2$ as the propagation angle $\theta$ increases from $\theta_c$ to 90$^\circ$. 
Here, $f_2$ is defined as 
\begin{equation}
f_2 = \sqrt{f_1^2+\left(\frac{{\gamma}\left(4{\pi}M_{\rm{eff}}-H_u\right)}{2}\right)^2}. 
\end{equation}
It should be noted that spin waves within a finite range of propagation angles $\theta$ were detected 
due to the finite sizes of the pump and probe spots in our experiment. 
Therefore, the detected range of $\theta$ depended on the scan length, which is defined as 
the distance between the pump and probe spots. 
This effect caused the spectral shape of MSWs to be dependent on the scan length, as will be mentioned below. 

Figure 3 is a representative spatial map of spin waves in the parallel geometry. 
The delay time was fixed to 2.1 ns. 
In this geometry, propagation of BVMSWs is expected \cite{Damon,Stancil}. 
In the shaded region, whose width corresponds to the sum of the spot diameters of the pump and probe, 
the thermal effect cannot be completely subtracted. 
Outside this region, propagation of spin waves with a wavelength $\lambda =$ 300 $\mu$m was observed. 
The spot sizes were confirmed to be small enough to resolve the waveform. 
The wavenumber ($k = 2\pi/\lambda$) also satisfies $k_0 \ll k \ll 1/l_0$.
Here, 
$k_0$ is the wavenumber of the electromagnetic waves that possess the same frequency as the spin waves, 
and $\l_0 = \sqrt{D/4{\pi}M_s}$ is an exchange length. 
Using a value of the exchange stiffness from the literature, $D = 9.4 \times 10^{-9}$ $\rm{Oe}$ $\rm{cm}^2$ \cite{Liu2}, 
$l_0$ is 29 nm. The exchange interaction 
is negligible when the above condition is satisfied. 
Therefore, these waves are regarded as being MSWs. 
The propagation in both directions was symmetric in amplitude and phase, 
unlike the excitation by a microwave field emitted from a microstrip antenna \cite{Schneider}. 
This is because the direction of the effective magnetic field by IFE is perpendicular to the sample surface, 
which affects the magnetization symmetrically about the pump spot. 

\begin{figure}[t]
\includegraphics[height=4cm,clip,keepaspectratio]{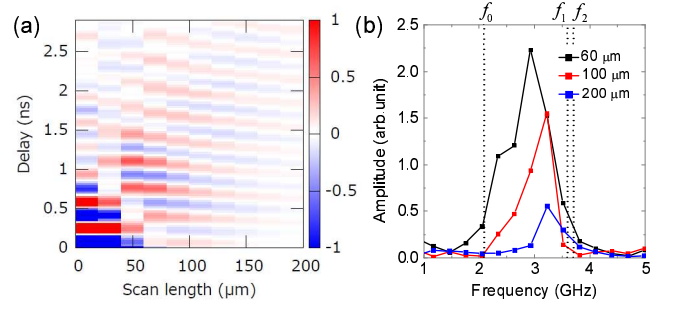}
\caption{(Color online) Spin waves when scanning perpendicular to the magnetic field. 
(a) A spatiotemporal map of spin waves. 
The color bar represents polarization rotation of the transmitted probe light in units of mrad.
(b) Dependence of Fourier spectrum of MSWs on scan length. 
The dotted lines respectively represent calculated upper and lower limits of the BVMSW band and the MSSW band.}
\label{FIG. 5.}
\end{figure}

Figure 4(a) shows a spatiotemporal map of the spin waves in the parallel geometry. 
It is clearly seen that a spin wave packet propagated from the pump spot, and 
the phase of the wave packet moved toward the pump position from the outside, which is a characteristic of BVMSWs. 
The dispersion of the lowest-order mode of the BVMSWs is approximated by \cite{Stancil} 
\begin{equation}
f\left(k\right) = {\gamma}\sqrt{H_0\left(H_0-H_u+4{\pi}M_{\rm{eff}}\frac{1-\exp(-kd)}{kd}\right)}.
\end{equation}
Here, $d$ = 110 $\mu$m is the thickness of the sample. 
When we use the experimentally determined wavenumber (2$\pi$/300 $\mu \rm{m}^{-1}$) of the MSWs, 
$f$ is calculated to be 2.8 GHz from Eq. (4), 
which is consistent with the experimentally determined frequency (2.9 GHz). 
The group velocity of the BVMSWs calculated from Eq. (4) is 135 km/s, 
which is shown as the dotted line in Fig. 4(a). 
This result is in good agreement with the experimental results. 
Figure 4(b) shows the dependence of the Fourier spectrum of the MSWs on the scan length in the parallel geometry. 
The upper limit $f_1$ and the lower limit $f_0$ of the BVMSW band, which are shown in Fig. 4(b), 
were calculated from Eqs. (1) and (2). 
The frequency range of the propagating MSWs, centered on 2.9 GHz, agreed well with the calculated BVMSW band 
(see Fig. 2). 
As the pump spot went away from the probe spot, only the propagation angle $\theta$ close to 0$^\circ$ was detected. 
The BVMSW band was the broadest for $\theta = 0^\circ$, in agreement with Fig. 2. 

Figure 5(a) shows a spatiotemporal map of the MSWs in the perpendicular geometry, 
where propagation of MSSWs was also expected in addition to BVMSWs. 
However, the phase of the wave packet went toward the pump position from the outside. 
This suggests that the BVMSWs were the dominant MSWs propagating in this direction. 
As the pump spot went away from the probe spot, only the propagation angle $\theta$ close to 90$^\circ$ 
was detected in this geometry. 
Because the BVMSW band width was reduced as the scan length increased, the group velocity of the BVMSWs became small 
in this direction. 
Unlike the parallel geometry, the wave packet stayed near the pump spot due to this effect. 
Phase discontinuity around the scan length of 40 $\mu$m is an artifact which is attributed to 
partial spatial overlap between the pump and probe spots. 
Figure 5(b) shows the dependence of the Fourier spectrum of the MSWs on the scan length in the perpendicular geometry. 
The spectral intensity at the lower frequency in the BVMSW band was suppressed more than that at the higher frequency 
as the scan length increased, because the lower limit of the BVMSW band moved up as the scan length increased. 
The upper limit of the MSSW band, $f_2$, in Fig. 5(b) was calculated from Eq. (3). 
The amplitude of the MSSWs was very small compared with that of the BVMSWs, 
which suggests that the MSSWs were not effectively excited by the pump pulses. 
Lenk \textit{et al}. \cite{Lenk} suggested that asymmetric excitation across the thickness of the sample is needed to excite MSSWs. 
Because of the negligible absorbance of the pump pulse, asymmetric excitation was not realized in our sample. 
Thus, the excitation efficiency of MSSWs was lower 
than that of BVMSWs, which were distributed almost uniformly across the thickness of the sample \cite{Patton}. 

In summary, two-dimensional propagation of photo-induced spin waves in Bi-doped 
rare-earth iron garnet was investigated by an all-optical pump-probe experiment. 
Propagating spin waves were impulsively excited by a circularly polarized light pulse via IFE. 
In the parallel geometry, BVMSWs were detected, and their frequency and wavelength were in good agreement with 
the theoretical dispersion relation. 
BVMSWs were also detected in the perpendicular geometry, 
with a higher frequency than that in the parallel geometry. 
This frequency dependence on the propagating direction is consistent with the theoretical prediction 
that the lower limit of the BVMSW band moves up as the propagation angle increases. 
MSSWs were not effectively excited, which can be attributed to the negligible absorbance of the pump pulse in our sample. 
Our findings open up the possibility of using IFE as an impulsive spin wave generator. 
This nonthermal, coherent excitation method will be an important tool in magnonics. 

We thank S. Takeda for the magnetic resonance measurements and valuable discussions. 
This work was supported by KAKENHI (23104706) and the JST PRESTO program.

\end{document}